\documentclass[italian,english]{article}
\usepackage[latin9]{inputenc}
\usepackage{color}
\usepackage{amssymb}
\usepackage{esint}

\newcommand{\lyxaddress}[1]{
\par {\raggedright #1
\vspace{1.4em}
\noindent\par}
}

\usepackage{babel}

\begin{document}

\title{\textbf{A clarification on the debate on {}``}\textbf{\emph{the
original Schwarzschild solution}}\textbf{''}}

\author{\textbf{Christian Corda }}

\maketitle

\lyxaddress{\begin{center}
International Institute for Theoretical Physics and Mathematics Einstein-Galilei,
via Santa Gonda 14, 59100 Prato, Italy and Institute for Basic Research,
P. O. Box 1577, Palm Harbor, FL 34682, USA .
\par\end{center}}

\begin{center}
\textit{E-mail addresses:} \textcolor{blue}{cordac.galilei@gmail.com}
\par\end{center}
\begin{abstract}
Now that English translations of Schwarzschild's original paper exist,
that paper has become accessible to more people. Historically, the
so-called \char`\"{}standard Schwarzschild solution\char`\"{} was
not the original Schwarzschild's work, but it is actually due to J.
Droste and, independently, H. Weyl, while it has been ultimately enabled
like correct solution by D. Hilbert. Based on this, there are authors
who claim that the work of Hilbert was wrong and that Hilbert's mistake
spawned black-holes and the community of theoretical physicists continues
to elaborate on this falsehood, with a hostile shouting down of any
and all voices challenging them. In this paper we re-analyse \char`\"{}the
original Schwarzschild solution\char`\"{} and we show that it is totally
equivalent to the solution enabled by Hilbert. Thus, the authors who
claim that \char`\"{}the original Schwarzschild solution\char`\"{}
implies the non existence of black holes give the wrong answer. We
realize that the misunderstanding is due to an erroneous interpretation
of the different coordinates. In fact, arches of circumference appear
to follow the law $dl=rd\varphi$, if the origin of the coordinate
system is a non-dimensional material point in the core of the black-hole,
while they do not appear to follow such a law, but to be deformed
by the presence of the mass of the central body $M$ if the origin
of the coordinate system is the surface of the Schwarzschild sphere. 
\end{abstract}


\section{Introduction}

The concept of black-hole (BH) has been considered very fascinating
by scientists even before the introduction of general relativity (see
\cite{key-1} for an historical review). A BH is a region of space
from which nothing, not even light, can escape. It is the result of
the deformation of spacetime caused by a very compact mass. Around
a BH there is an undetectable surface which marks the point of no
return. This surface is called an event horizon. It is called \char`\"{}black\char`\"{}
because it absorbs all the light that hits it, reflecting nothing,
just like a perfect black body in thermodynamics \cite{key-2}. However,
an unsolved problem concerning such objects is the presence of a space-time
singularity in their core. Such a problem was present starting by
the firsts historical papers concerning BHs \cite{key-3,key-4,key-5}.
It is a common opinion that this problem could be solved when a correct
quantum gravity theory will be, finally, obtained, see \cite{key-6}
for recent developments.

On the other hand, fundamental issues which dominate the question
about the existence or non-existence of BH horizons and singularities
and some ways to avoid the development of BH singularities within
the classical theory, which does not require the need for a quantum
gravity theory, have been discussed by various authors in the literature,
see references from \cite{key-7} to \cite{key-16}. In fact, by considering
the exotic nature of BHs, it may be natural to question if such bizarre
objects could exist in nature or to suggest that they are merely pathological
solutions to Einstein's equations. Einstein himself thought that BHs
would not form, because he held that the angular momentum of collapsing
particles would stabilize their motion at some radius \cite{key-17}.

Recently, the debate became very hot as English translations of Schwarzschild's
original work now exist and that work has become accessible to more
people \cite{key-18,key-19}. Historically, the so-called \char`\"{}\emph{Schwarzschild
solution}'' was not the original Schwarzschild's work, but it is
actually due to J. Droste \cite{key-20} and, independently, H. Weyl
\cite{key-21}. while it has been ultimately enabled like correct
solution by D. Hilbert \cite{key-22}. Let us further clarify this
point by adding some historical notes. In 1915, A. Einstein developed
his theory of general relativity \cite{key-23}. A few months later,
K. Schwarzschild gave the solution for the gravitational field of
a point mass and a spherical mass \cite{key-3}. A few months after
Schwarzschild, J. Droste, a student of H. Lorentz, independently gave
an apparently different solution for the point mass and wrote more
extensively about its properties \cite{key-20}. In such a work Droste
also claimed that his solution was physically equivalent to the one
by Schwarzschild. In the same year, 1917, H. Weyl re-obtained the
same solution by Droste \cite{key-21}. This solution had a peculiar
behaviour at what is now called the Schwarzschild radius, where it
became singular, meaning that some of the terms in the Einstein equations
became infinite. The nature of this surface was not quite understood
at the time, but Hilbert \cite{key-22} claimed that the form by Droste
and Weyl was preferable to that in \cite{key-3} and ever since then
the phrase \textquotedblleft{}\emph{Schwarzschild solution}\textquotedblright{}
has been taken to mean the line-element which was found in \cite{key-20,key-21}
rather than the original solution in \cite{key-2}. In 1924, A. Eddington
showed that the singularity disappeared after a change of coordinates
(Eddington coordinates \cite{key-24}), although it took until 1933
for G. Lemaître to realize, in a series of lectures together with
Einstein, that this meant the singularity at the Schwarzschild radius
was an unphysical coordinate singularity \cite{key-25}.

In 1931, S. Chandrasekhar calculated that a non-rotating body of electron-degenerate
matter above 1.44 solar masses (the Chandrasekhar limit) would collapse
\cite{key-6}. His arguments were opposed by many of his contemporaries
like Eddington, Lev Landau and the same Einstein. In fact, a white
dwarf slightly more massive than the Chandrasekhar limit will collapse
into a neutron star which is itself stable because of the Pauli exclusion
principle \cite{key-1}. But in 1939, \foreignlanguage{italian}{J.
R. Oppenheimer and G. M. Volkoff} predicted that neutron stars above
approximately 1.5 - 3 solar masses (the famous Oppenheimer\textendash{}Volkoff
limit) would collapse into BHs for the reasons presented by Chandrasekhar,
and concluded that no law of physics was likely to intervene and stop
at least some stars from collapsing to BHs \cite{key-26}. Oppenheimer
and \foreignlanguage{italian}{Volkoff} interpreted the singularity
at the boundary of the Schwarzschild radius as indicating that this
was the boundary of a bubble in which time stopped. This is a valid
point of view for external observers, but not for free-falling observers.
Because of this property, the collapsed stars were called \char`\"{}\emph{frozen
stars}\char`\"{} \cite{key-27} because an outside observer would
see the surface of the star frozen in time at the instant where its
collapse takes it inside the Schwarzschild radius. This is a known
property of modern BHs, but it must be emphasized that the light from
the surface of the frozen star becomes redshifted very fast, turning
the BH black very quickly. Originally, many physicists did not accept
the idea of time standing still at the Schwarzschild radius, and there
was little interest in the subject for lots of time. But in 1958,
D. Finkelstein, by re-analysing Eddington coordinates, identified
the Schwarzschild surface $r=2M$ (in \emph{natural units}, i.e. $G=1$,
$c=1$ and $\hbar=1$, i.e where $r$ is the radius of the surface
and $M$ is the mass of the BH) as an \emph{event horizon}, \char`\"{}\emph{a
perfect unidirectional membrane: causal influences can cross it in
only one direction}\char`\"{} \cite{key-28}. This extended Oppenheimer's
results in order to include the point of view of free-falling observers.
Finkelstein's solution extended the Schwarzschild solution for the
future of observers falling into the BH. Another complete extension
was found by M. Kruskal in 1960 \cite{key-29}. 

These results generated a new interest on general relativity, which,
together with BHs, became mainstream subjects of research within the
Scientific Community. This process was endorsed by the discovery of
pulsars in 1968 \cite{key-30} which resulted to be rapidly rotating
neutron stars. Until that time, neutron stars, like BHs, were regarded
as just theoretical curiosities; but the discovery of pulsars showed
their physical relevance and spurred a further interest in all types
of compact objects that might be formed by gravitational collapse.

In this period more general BH solutions were found. In 1963, R. Kerr
found the exact solution for a rotating BH \cite{key-31}. Two years
later E. T. Newman and A. Janis found the asymmetric solution for
a BH which is both rotating and electrically charged \cite{key-32}.
Through the works by W. Israel, B. Carter and D. C. Robinson the no-hair
theorem emerged \cite{key-1}, stating that a stationary BH solution
is completely described by the three parameters of the Kerr\textendash{}Newman
metric; mass, angular momentum, and electric charge \cite{key-1}.

For a long time, it was suspected that the strange features of the
BH solutions were pathological artefacts from the symmetry conditions
imposed, and that the singularities would not appear in generic situations.
This view was held in particular by Belinsky, Khalatnikov, and Lifshitz,
who tried to prove that no singularities appear in generic solutions
\cite{key-1}. However, in the late sixties R. Penrose and S. Hawking
used global techniques to prove that singularities are generic \cite{key-1}.

The term \char`\"{}\emph{black hole}\char`\"{} was first publicly
used by J. A. Wheeler during a lecture in 1967 \cite{key-33} but
the first appearing of the term, in 1964, is due by A. Ewing in a
letter to the American Association for the Advancement of Science
\cite{key-34}, verbatim: {}``\emph{According to Einstein\textquoteright{}s
general theory of relativity, as mass is added to a degenerate star
a sudden collapse will take place and the intense gravitational field
of the star will close in on itself. Such a star then forms a \textquoteleft{}black
hole\textquoteright{} in the universe}.'' 

In any case, after Wheeler's use of the term, it was quickly adopted
in general use. 

Today, the majority of researchers in the field is persuaded that
there is no obstacle to forming an event horizon. On the other hand,
there are other researchers who demonstrated that various physical
mechanisms can, in principle, remove both of event horizon and singularities
during the gravitational collapse \cite{key-7} - \cite{key-16}.
In particular, in \cite{key-9} an exact solution of Einstein field
equations which removes both of event horizon and singularities has
been found by constructing the right-hand side of the field equations,
i.e. the stress-energy tensor, through a non-linear electrodynamics
Lagrangian which was previous used in super-strongly magnetized compact
objects, such as pulsars, and particular neutron stars \cite{key-35,key-36}.

On the other hand, there are researchers who invoke the non existence
of BH by claiming that the Schwarzschild's original work \cite{key-3}
gives a solution which is physically different from the one derived
by Droste \cite{key-20} and Weyl \cite{key-21}. Let us see this
issue in more detail. The new translations of Schwarzschild's original
work can be found in ref. \cite{key-18,key-19}. These works commented
on Schwarzschild's original paper \cite{key-3}. In particular Abrams
\cite{key-18} claimed that the line-element (we use natural units
in all this paper)

\begin{equation}
ds^{2}=(1-\frac{r_{g}}{r})dt^{2}-r^{2}(\sin^{2}\theta d\varphi^{2}+d\theta^{2})-\frac{dr^{2}}{1-\frac{r_{g}}{r}}\label{eq: Hilbert}\end{equation}
 i.e. the famous and fundamental solution to the Einstein field equations
in vacuum, gives rise to a space-time that is neither equivalent to
Schwarzschild's original solution in \cite{key-3}. Abrams also claimed
that Hilbert \cite{key-22} opined that the form of (\ref{eq: Hilbert})
by Droste and Weyl was preferable to that in \cite{key-3} and ever
since then the phrase \textquotedblleft{}\emph{Schwarzschild solution}\textquotedblright{}
has been taken to mean the line-element (\ref{eq: Hilbert}) rather
than the original solution in \cite{key-3}. In a following work \cite{key-37}
Abrams further claimed that {}``\emph{Black Holes are The Legacy
of Hilbert's Error}'' as Hilbert's derivation used a wrong variable.
Thus, Hilbert's assertion that the form of (\ref{eq: Hilbert}) was
preferable to the original one in \cite{key-3} should be invalid.
Based on this, there are authors who agree with Abrams by claiming
that the work of Hilbert was wrong and Hilbert's mistake spawned the
BHs and the community of theoretical physicists continues to elaborate
on this falsehood, with a hostile shouting down of any and all voices
challenging them, see for example references \cite{key-38,key-39,key-40,key-41}.

In this paper we re-analyse \emph{{}``the original Schwarzschild
solution''} to Einstein field equations derived in \cite{key-3}.
Such a solution arises from an apparent different physical hypothesis
which assumes arches of circumference to to \emph{do not follow the
law }$dl=rd\varphi$, but to be deformed by the presence of the mass
of the central body $M.$ This assumption enables the origin of the
coordinate system to be \emph{not} a single point, but a spherical
surface having radius equal to the gravitational radius, i.e. the
surface of the Schwarzschild sphere. The solution works for the external
geometry of a spherical static star and circumnavigates the Birkhoff
theorem \cite{key-4}. 

Then, the simplest case of gravitational collapse, i.e. the spherical
radial collapse of a star with uniform density and zero pressure,
will be analysed by turning attention to the interior of the collapsing
object and the precise word line that its surface follows in the external
geometry. The result of the analysis will show that the singularity
within the totally collapsed spherical object remains. In fact, a
coordinate transform that transfers the origin of the coordinate system,
which is the surface of a sphere having radius equal to the gravitational
radius, in a non-dimensional material point in the core of the BH
re-obtains the solution (\ref{eq: Hilbert}). Thus, \emph{{}``the
original Schwarzschild solution''} \cite{key-3} results physically
equivalent to the solution (\ref{eq: Hilbert}) enabled like the correct
on by Hilbert in \cite{key-23}, i.e. the solution that is universally
known like the \char`\"{}Schwarzschild solution\char`\"{} \cite{key-1}.
This analysis ultimately shows that the authors who claim that the
original Schwarzschild solution leaves no room for the science fiction
of the BHs (see references\cite{key-18,key-19} and from \cite{key-37}
to \cite{key-41}) give the wrong answer. The misunderstanding is
due to an erroneous interpretation of the different coordinates. In
fact, arches of circumference appear to follow the law\emph{ }$dl=rd\varphi$,
if the origin of the coordinate system is a non-dimensional material
point in the core of the BH, while they do not appear to follow such
a law, but to be deformed by the presence of the mass of the central
body $M$ if the origin of the coordinate system is the surface of
the Schwarzschild sphere. Thus, the only way to remove the singularity
in the core of a BH within the classical theory of Einstein's general
relativity is changing the hypotheses which govern the internal geometry
of the collapsing star, following for example the ideas in references
from \cite{key-7} to \cite{key-16}.

\section{The\emph{ {}``original Schwarzschild solution''}}

Following \cite{key-42}, the more general line-element which respects
central symmetry is

\begin{equation}
ds^{2}=h(r,t)dr^{2}+k(r,t)(\sin^{2}\theta d\varphi^{2}+d\theta^{2})+l(r,t)dt^{2}+a(r,t)drdt,\label{eq: generale}\end{equation}

where \begin{equation}
\begin{array}{ccccc}
r\geq0, &  & 0\leq\theta\leq\pi, &  & 0\leq\varphi\leq2\pi.\end{array}\label{eq: assunzioni}\end{equation}

We search a line-element solution in which the metric is spatially
symmetric with respect to the origin of the coordinate system, i.e.
that we find again the same solution when spatial coordinates are
subjected to a orthogonal transformations and rotations and it is
asymptotically flat at infinity \cite{key-1}. In order to obtain
the {}``\emph{standard Schwarzschild solution}'', i.e. the line-element
(\ref{eq: Hilbert}) to Einstein field equations in vacuum one uses
transformations of the type \cite{key-42}

\begin{equation}
\begin{array}{ccc}
r=f_{1}(r',t'), &  & t=f_{2}(r',t')\end{array},\label{eq: transformations}\end{equation}

where $f_{1}$ and $f_{1}$ are arbitrary functions of the new coordinates
$r'$ and $t'$. At this point, if one wants the {}``\emph{standard
Schwarzschild solution}'', $r$ and $t$ have to be chosen in a way
that $a(r,t)=0$ and $k(r,t)=-r^{2}$ \cite{key-42}. In particular,
the second condition implies that the standard Schwarzschild radius
is determined in a way which guarantees that the length of the circumference
centred in the origin of the coordinate system is $2\pi r$ \cite{key-42}.

In our approach, we will suppose again that $a(r,t)=0,$ but, differently
from the standard analysis, we will assume that the length of the
circumference centred in the origin of the coordinate system \emph{is
not} $2\pi r$. We release \emph{an apparent different physical assumption},
i.e. that arches of circumference \emph{are deformed} by the presence
of the mass of the central body $M.$ Note that this different physical
hypothesis permits to circumnavigate the Birkhoff Theorem \cite{key-4}
which leads to the {}``\emph{standard Schwarzschild solution}''
\cite{key-3}. In fact, the demonstration of the Birkhoff Theorem
\emph{starts} from a line element \emph{in which $k(r,t)=-r^{2}$
has been chosen}, see the discussion in paragraph 32.2 of \cite{key-1}
and, in particular, look at Eq. (32.2) of such a paragraph.

Then, we proceed assuming $k=-mr^{2},$ where $m$ is a generic function
to be determined in order to obtain that the length of circumferences
centred in the origin of the coordinate system\emph{ }are not $2\pi r$.
In other words, $m$ represents \emph{a measure of the deviation from
}$2\pi r$ of circumferences centred in the origin of the coordinate
system.

The line element (\ref{eq: generale}) becomes

\begin{equation}
ds^{2}=hdr^{2}-mr^{2}(\sin^{2}\theta d\varphi^{2}+d\theta^{2})+ldt^{2}.\label{eq: generale 2}\end{equation}

One puts \begin{equation}
\begin{array}{c}
X\equiv\frac{1}{3}r^{3}\\
\\Y\equiv-\cos\theta\\
\\Z\equiv\varphi.\end{array}\label{eq: trasforma}\end{equation}

In the $X,Y,Z$ coordinates the line-element (\ref{eq: generale 2})
reads

\begin{equation}
ds^{2}=ldt^{2}+\frac{h}{^{r^{4}}}dX^{2}-mr^{2}[\frac{dY^{2}}{1-Y^{2}}+dZ^{2}(1-Y^{2})].\label{eq: alternative}\end{equation}

Let us consider three functions 

\label{eq: nuove funzioni}\begin{equation}
\begin{array}{c}
A\equiv-\frac{h}{^{r^{4}}}\\
\\B\equiv mr^{2}\\
\\C\equiv l\end{array}\label{eq: functions}\end{equation}

which satisfy the conditions \begin{equation}
\begin{array}{c}
X\rightarrow\infty\mbox{ }implies\mbox{ }A\rightarrow\frac{1}{r^{4}}=\frac{1}{(3X)^{\frac{4}{3}}},\mbox{ }B\rightarrow r^{2}=3X^{\frac{2}{3}},\mbox{ }C\rightarrow1\\
\\normalization\mbox{ }condition\mbox{ }AB^{2}C=1.\end{array}\label{eq: condizioni}\end{equation}

The line-element (\ref{eq: alternative}) becomes

\begin{equation}
ds^{2}=Cdt^{2}-AdX^{2}-B\frac{dY^{2}}{1-Y^{2}}-BdZ^{2}(1-Y^{2}).\label{eq: alternative 2}\end{equation}

From the metric (\ref{eq: alternative 2}) one gets the Christoffell
coefficients like (only the non zero elements will be written down)

\begin{equation}
\begin{array}{ccc}
\Gamma_{tX}^{t}=-\frac{1}{2C}\frac{\partial C}{\partial X} &  & \Gamma_{XX}^{X}=-\frac{1}{2A}\frac{\partial A}{\partial X}\\
\\\Gamma_{YY}^{X}=\frac{1}{2A}\frac{\partial B}{\partial X}\frac{1}{1-Y^{2}} &  & \Gamma_{ZZ}^{X}=\frac{1}{2A}\frac{\partial B}{\partial X}(1-Y^{2})\\
\\\Gamma_{tt}^{X}=-\frac{1}{2A}\frac{\partial C}{\partial X} &  & \Gamma_{YX}^{Y}=-\frac{1}{2B}\frac{\partial B}{\partial X}\\
\\\Gamma_{YY}^{Y}=\frac{-Y}{1-Y^{2}} &  & \Gamma_{ZZ}^{Y}=-Y(1-Y^{2})\\
\\\Gamma_{ZX}^{Z}=-\frac{1}{2B}\frac{\partial B}{\partial X} &  & \Gamma_{ZX}^{Z}=\frac{Y}{1-Y^{2}}.\\
\\\end{array}\label{eq: connessioni}\end{equation}

By using the equation for the components of the Ricci tensor, the
components of Einstein field equation in vacuum are \cite{key-42}

\begin{equation}
R_{ik}=\frac{\partial\Gamma_{ik}^{l}}{dx^{l}}-\frac{\partial\Gamma_{il}^{l}}{dx^{k}}-\Gamma_{ik}^{l}\Gamma_{lm}^{m}-\Gamma_{il}^{m}\Gamma_{km}^{l}=0.\label{eq: ricci}\end{equation}

By inserting Eqs. (\ref{eq: connessioni}) in Eqs. (\ref{eq: ricci})
one gets only three independent relations

\begin{equation}
\frac{\partial}{\partial X}(\frac{1}{A}\frac{\partial B}{\partial X})-2-\frac{1}{AB}(\frac{\partial B}{\partial X})^{2}=0\label{eq: einstein 2}\end{equation}
\begin{equation}
\frac{\partial}{\partial X}(\frac{1}{A}\frac{\partial A}{\partial X})-\frac{1}{2}(\frac{1}{A}\frac{\partial A}{\partial X})^{2}-(\frac{1}{B}\frac{\partial B}{\partial X})^{2}-\frac{1}{2}(\frac{1}{C}\frac{\partial C}{\partial X})^{2}=0\label{eq: einstein 1}\end{equation}

\begin{equation}
\frac{\partial}{\partial X}(\frac{1}{A}\frac{\partial C}{\partial X})-\frac{1}{AC}(\frac{\partial C}{\partial X})^{2}=0.\label{eq: einstein 3}\end{equation}

From the second of Eqs. (\ref{eq: condizioni}) (normalization condition)
one gets also

\begin{equation}
\frac{1}{A}\frac{\partial A}{\partial X}+\frac{2}{B}\frac{\partial B}{\partial X}+\frac{1}{C}\frac{\partial C}{\partial X}=0.\label{eq: normalizzata}\end{equation}

Eq. (\ref{eq: einstein 3}) can be rewritten like

\begin{equation}
\frac{\partial}{\partial X}(\frac{1}{C}\frac{\partial C}{\partial X})=\frac{1}{AC}\frac{\partial A}{\partial X}\frac{\partial C}{\partial X},\label{eq: einstein 3.1}\end{equation}

which can be integrated, giving 

\begin{equation}
\frac{1}{C}\frac{\partial C}{\partial X}=aA,\label{eq: integra 1}\end{equation}

where $a$ is an integration constant.

By adding Eq. (\ref{eq: einstein 1}) to Eq. (\ref{eq: einstein 3.1})
one gets

\begin{equation}
\frac{\partial}{\partial X}(\frac{1}{A}\frac{\partial A}{\partial X}+\frac{1}{C}\frac{\partial C}{\partial X})=(\frac{1}{B}\frac{\partial B}{\partial X})^{2}+\frac{1}{2}(\frac{1}{A}\frac{\partial A}{\partial X}+\frac{1}{C}\frac{\partial C}{\partial X})^{2}\label{eq: einstein 1.1}\end{equation}

Considering Eq. (\ref{eq: normalizzata}) we obtain

\begin{equation}
2\frac{\partial}{\partial X}(\frac{1}{B}\frac{\partial B}{\partial X})=-3(\frac{1}{B}\frac{\partial B}{\partial X})^{2},\label{eq: einstein 2.1}\end{equation}

which can be integrated, giving

\begin{equation}
\frac{1}{B}\frac{\partial B}{\partial X}=\frac{2}{3X+b},\label{eq: integra 2}\end{equation}

where $b$ is an integration constant. A second integration gives

\begin{equation}
B=d(3X+b)^{\frac{2}{3}}.\label{eq: integra 3}\end{equation}

where $d$ is an integration constant. But the first of Eqs. (\ref{eq: condizioni})
implies $d=1,$ thus \begin{equation}
B=(3X+b)^{\frac{2}{3}}.\label{eq: integrata}\end{equation}

By using Eqs. (\ref{eq: integra 1}) and (\ref{eq: normalizzata})
we obtain \[
\frac{\partial C}{\partial X}=aAC=\frac{a}{B^{2}}=a(3X+b)^{-\frac{4}{3}}.\]

By integrating and considering the first of Eqs. (\ref{eq: condizioni})
one gets

\begin{equation}
C=1-a(3X+b)^{-\frac{1}{3}}\label{eq: C}\end{equation}

Then, from Eq. (\ref{eq: normalizzata}) one obtains

\begin{equation}
A=\frac{(3X+b)^{-\frac{4}{3}}}{1-a(3X+b)^{-\frac{1}{3}}}.\label{eq: singular}\end{equation}

By putting Eqs. (\ref{eq: singular}) and (\ref{eq: integrata}) in
Eq. (\ref{eq: einstein 2}) one immediately sees that this last equation
is automatically satisfied.

We note that the function $A$ results singular for values $a(3X+b)^{-\frac{1}{3}}=1.$
However, this is a mathematical singularity due to the particular
coordinates $t,X,Y,Z$ defined by the transformation (\ref{eq: trasforma}).
In fact, by assuming that such a singularity is located at $X=0$
we get

\begin{equation}
b=a^{3},\label{eq: ba}\end{equation}

i.e. we find a relation between the two integration constants $b$
and $a$. 

At the end we obtain

\begin{equation}
\begin{array}{c}
A=(r^{3}+a^{3})^{-\frac{4}{3}}[1-a(r^{3}+a^{3})^{-\frac{1}{3}}]^{-1}\\
\\B=(r^{3}+a^{3})^{\frac{2}{3}}\\
\\C=1-a(r^{3}+a^{3})^{-\frac{1}{3}}.\end{array}\label{eq: soluzioni}\end{equation}

By inserting the functions (\ref{eq: soluzioni}) in Eq. (\ref{eq: alternative 2})
and using Eqs. (\ref{eq: functions}) and (\ref{eq: trasforma}) to
return to the standard polar coordinates the line-element solution
reads 

\begin{equation}
\begin{array}{c}
ds^{2}=\left[1-\frac{a}{(r^{3}+a^{3})^{\frac{1}{3}}}\right]dt^{2}-(r^{3}+a^{3})^{\frac{2}{3}}(\sin^{2}\theta d\varphi^{2}+d\theta^{2})+\\
\\-\frac{d(r^{3}+a^{3})^{\frac{2}{3}}}{1-\frac{a}{(r^{3}+a^{3})^{\frac{1}{3}}}}.\end{array}\label{eq: sol. generale}\end{equation}

Hence, we understand that the assumption to locate the mathematical
singularity of the function $A$ at $X=0$ coincides with the physical
condition that the length of the circumference centred in the origin
of the coordinate system is $2\pi(r^{3}+a^{3})^{\frac{1}{3}}$, which
is different from the value $2\pi r.$ This is the apparent fundamental
physical difference between this solution and the {}``\emph{standard
Schwarzschild solution}'' (\ref{eq: Hilbert}), i.e. the one enabled
by Hilbert in \cite{key-22}. The value of the generic function $m$
which permits that the length of circumferences centred in the origin
of the coordinate system\emph{ }are not $2\pi r$ is

\begin{equation}
m=\frac{(r^{3}+a^{3})^{\frac{2}{3}}}{r^{2}}.\label{eq: funzione m}\end{equation}

On the other hand, in order to determinate the value of the constant
$a$, by following \cite{key-42}, one can use the weak field approximation
which implies $g_{00}\cong1+2\varphi$ at large distances, where $g_{00}=(1-\frac{a}{(r^{3}+a^{3})^{\frac{1}{3}}})$
in Eq. (\ref{eq: sol. generale}) and $\varphi\equiv\frac{-M}{r}$
is the Newtonian potential. Thus, for $a\ll r,$ we immediately obtain:
$a=2M=r_{g},$ i.e. $a$ results exactly the gravitational radius
\cite{key-1,key-28}. 

Then, we can rewrite the solution (\ref{eq: sol. generale}) in an
ultimate way like

\begin{equation}
\begin{array}{c}
ds^{2}=\left[1-\frac{r_{g}}{(r^{3}+r_{g}^{3})^{\frac{1}{3}}}\right]dt^{2}-(r^{3}+r_{g}^{3})^{\frac{2}{3}}(\sin^{2}\theta d\varphi^{2}+d\theta^{2})+\\
\\-\frac{d(r^{3}+r_{g}^{3})^{\frac{2}{3}}}{1-\frac{r_{g}}{(r^{3}+r_{g}^{3})^{\frac{1}{3}}}}.\end{array}\label{eq: sol. gen.fin.}\end{equation}

Historically, the line-element (\ref{eq: sol. gen.fin.}) represents
\emph{{}``the} \emph{original Schwarzschild solution''} to Einstein
field equations as it has been derived for the first time by Karl
Schwarzschild in \cite{key-3} with a slight different analysis.

Some comments are needed. By looking Eq. (\ref{eq: sol. gen.fin.})
one understands that the origin of the coordinate that we have chosen
by putting $r\geq0,$ $0\leq\theta\leq\pi$ $0\leq\varphi\leq2\pi$
and with the additional assumption\emph{ }that the length of circumferences
centred in the origin of the coordinate system\emph{ }are not Euclidean,\emph{
is not} a single point, but it is the surface of a sphere having radius
$r_{g}$, i.e. the surface of the Schwarzschild sphere. By putting 

\begin{equation}
\widehat{r}\equiv(r^{3}+r_{g}^{3})^{\frac{1}{3}},\label{eq: definizione}\end{equation}

Eq. (\ref{eq: sol. generale}) becomes

\begin{equation}
ds^{2}=(1-\frac{r_{g}}{\widehat{r}})dt^{2}-\widehat{r}^{2}(\sin^{2}\theta d\varphi^{2}+d\theta^{2})-\frac{d\widehat{r}^{2}}{1-\frac{r_{g}}{\widehat{r}}}.\label{eq: sol. gen 2}\end{equation}
Eq. (\ref{eq: sol. gen 2}) looks \emph{formally} \emph{equal} to
the \emph{{}``standard Schwarzschild solution}'' (\ref{eq: Hilbert}).
But one could think that the transformation (\ref{eq: definizione})
is forbidden for the following motivation. It transfers the origin
of the coordinate system, $r=0,$ $\theta=0,$ $\varphi=0,$ which
is the surface of a sphere having radius $r_{g}$ in the $r,\mbox{ }\theta,\mbox{ }\varphi$
coordinates, in a non-dimensional material point $\widehat{r}=0,$
$\theta=0,$ $\varphi=0$ in the $\widehat{r},\mbox{ }\theta,\mbox{ }\varphi$
coordinates. Such a non-dimensional material point corresponds to
the point $r=-r_{g},$ $\theta=0,$ $\varphi=0$ in the original $r,\mbox{ }\theta,\mbox{ }\varphi$
coordinates. Thus, the transformation (\ref{eq: definizione}) could
not be a suitable coordinate transformation because it transfers a
spherical surface, i.e. a bi-dimensional manifold, in a non-dimensional
material point. We will see in the following that this interpretation
\emph{is not correct}.

On the other hand, we are searching a solution for the external geometry,
thus we assumed $r\geq0$ in Eq. (\ref{eq: assunzioni}) and from
Eq. (\ref{eq: definizione}) it is \emph{always} $\widehat{r}\geq r_{g}$
in Eq. (\ref{eq: sol. gen 2}) as it is $r\geq0$ in Eq. (\ref{eq: sol. gen.fin.}).
In this way, \emph{there are not physical singularities} in Eq. (\ref{eq: sol. gen 2}).
In fact, $r=0$ in Eq. (\ref{eq: sol. gen.fin.}) implies $\widehat{r}=r_{g}$
in Eq. (\ref{eq: sol. gen 2}) which corresponds to the mathematical
singularity at $X=0.$ This singularity is \emph{not physical} but
is due to the particular coordinates $t,X,Y,Z$ defined by the transformation
(\ref{eq: trasforma}).

Again, we emphasize the apparent different assumption of our analysis.\emph{
}As it is carefully explained in \cite{key-42}, the \emph{{}``standard
Schwarzschild solution}'' (\ref{eq: Hilbert}), arises from the hypothesis
that the coordinates $r$ and $t$ of the two functions (\ref{eq: transformations})
are chosen in order to guarantee that the length of the circumference
centred in the origin of the coordinate system is $2\pi r$. Indeed,
in the above derivation of \emph{{}``the} \emph{original Schwarzschild
solution}'' (\ref{eq: sol. gen.fin.}), $r$ and $t$ are chosen
in order to guarantee that the length of the circumference centred
in the origin of the coordinate system is \emph{not} $2\pi r$. In
particular, choosing to put the mathematical singularity of the function
$A$ at $X=0$ is equivalent to the physical condition that the length
of the circumference centred in the origin of the coordinate system
is $2\pi(r^{3}+r_{g}^{3})^{\frac{1}{3}}$. Then, one could think that
by forcing the transformation (\ref{eq: definizione}) for $r\leq0$,
one returns to the standard Schwarzschild solution (\ref{eq: Hilbert}),
but a bi-dimensional spherical surface, that is the surface of the
Schwarzschild sphere, is forced to become a non-dimensional material
point and we force a non-Euclidean geometry for circumferences to
become Euclidean. In that case, such a mathematical forcing could
be the cause of the singularity in the core of the black-hole. Thus,
this singularity could be only mathematical and not physical. But
in the following, by matching with the internal geometry, we will
see that this interpretation is not correct and that the singularity
in the core of the BH remains a physical singularity also in the case
of the \emph{{}``original Schwarzschild solution}'' given by Eq.
(\ref{eq: sol. gen.fin.}).

Notice that a large distances, i.e. where $r_{g}\ll r,$ the solution
(\ref{eq: sol. gen.fin.}) well approximates the standard Schwarzschild
solution (\ref{eq: Hilbert}), thus, both of the weak field approximation
and the analysis of astrophysical situations remain the same.

\section{Matching with the internal geometry: singular gravitational collapse}

In the following we adapt the classical analysis in \cite{key-1}
to the line-element (\ref{eq: sol. gen.fin.}). Let us consider a
test particle moving in the external geometry (\ref{eq: sol. gen.fin.}).
By following the magnitude of the 4-vector of energy-momentum is represented
by the rest mass $\mu$ of the particle \cite{key-1}

\begin{equation}
g_{ik}p^{i}p^{k}+\mu^{2}=g^{ik}p_{i}p_{k}+\mu^{2}=0,\label{eq: nullo}\end{equation}

or 

\begin{equation}
-\frac{E^{2}}{1-\frac{r_{g}}{(r^{3}+r_{g}^{3})^{\frac{1}{3}}}}+\frac{1}{1-\frac{r_{g}}{(r^{3}+r_{g}^{3})^{\frac{1}{3}}}}\frac{r^{4}}{(r^{3}+r_{g}^{3})^{\frac{4}{3}}}\left(\frac{dr}{d\lambda}\right)^{2}+\frac{L^{2}}{(r^{3}+r_{g}^{3})^{\frac{2}{3}}}+\mu^{2},\label{eq: casino}\end{equation}

where $\lambda=\tau/\mu$, $L$ and $E$ represent the affine parameter
being $\tau$ the proper time, the angular momentum and the energy
of the particle \cite{key-1}.

Einstein equivalence principle \cite{key-1} implies that test particles
follow the same wordlines regardless of mass. Then, what is relevant
for the motion of particles are the normalized quantities $\tilde{L}=L/\mu$
and $\tilde{E}=E/\mu$.

Thus, Eq. (\ref{eq: casino}) can be rewritten as

\begin{equation}
\begin{array}{c}
\left(\frac{dr}{d\tau}\right)^{2}=\frac{(r^{3}+r_{g}^{3})^{\frac{4}{3}}}{r^{4}}\left\{ \tilde{E}^{2}-\left(1-\frac{r_{g}}{(r^{3}+r_{g}^{3})^{\frac{1}{3}}}\right)\left(1+\frac{\tilde{L}^{2}}{(r^{3}+r_{g}^{3})^{\frac{2}{3}}}\right)\right\} =\\
\\=\frac{(r^{3}+r_{g}^{3})^{\frac{4}{3}}}{r^{4}}\left(\tilde{E}^{2}-\tilde{V}^{2}(r)\right),\end{array}\label{eq: casino 2}\end{equation}

where the {}``\emph{effective potential}'' is defined by 

\begin{equation}
\tilde{V}(r)\equiv\sqrt{\left(1-\frac{r_{g}}{(r^{3}+r_{g}^{3})^{\frac{1}{3}}}\right)\left(1+\frac{\tilde{L}^{2}}{(r^{3}+r_{g}^{3})^{\frac{2}{3}}}\right)}.\label{eq: effettivo}\end{equation}

From Eq. (\ref{eq: casino 2}) the proper time can be explicitly written
down \begin{equation}
\tau=\int d\tau=\int\left[\frac{dr}{\left(\tilde{E}^{2}-\tilde{V}^{2}(r)\right)}\frac{r^{2}}{(r^{3}+r_{g}^{3})^{\frac{2}{3}}}\right].\label{eq: tau}\end{equation}

In the following we discuss the collapse of a star with uniform density
and zero pressure. Because no pressure gradients are present to deflect
their motion, the particles on the surface of any ball of dust must
move along radial geodesic in the external geometry of Eq. (\ref{eq: sol. gen.fin.}).
The angular momentum vanishes and the integral (\ref{eq: tau}) reduces
to

\begin{equation}
\tau=\int d\tau=\int\left[\frac{dr}{\sqrt{\left(\frac{r_{g}}{(r^{3}+r_{g}^{3})^{\frac{1}{3}}}-\frac{r_{g}}{(R^{3}+r_{g}^{3})^{\frac{1}{3}}}\right)}}\frac{r^{2}}{(r^{3}+r_{g}^{3})^{\frac{2}{3}}}\right],\label{eq: tau 2}\end{equation}

where $R\equiv\frac{r_{g}}{1-\tilde{E}^{2}}$ is the \emph{{}``apastron}'',
i.e. the radius at which the particle has zero velocity \cite{key-1}.

Eq. (\ref{eq: tau 2}) can be integrated in parametric form:

\begin{equation}
r=\frac{1}{2}\left[(R^{3}+r_{g}^{3})(1+\cos\eta)^{3}-r_{g}^{3}\right]^{\frac{1}{3}}\label{eq: r}\end{equation}

and

\begin{equation}
\tau=\frac{(R^{3}+r_{g}^{3})^{\frac{1}{3}}}{2}\left(\frac{R^{3}+r_{g}^{3}}{r_{g}}\right)^{\frac{1}{6}}(\eta+\sin\eta).\label{eq: tau finale}\end{equation}

Eq. (\ref{eq: tau finale}) is the proper time read by a clock on
the surface of the collapsing star.

The collapse begins when the parameter $\eta$ is zero ($r=R$, $\tau=0$)
and terminates, for the external geometry, at $r=0$, $\eta=\frac{2r_{g}}{(R^{3}+r_{g}^{3})^{\frac{1}{3}}-1}.$ 

Thus, the total proper time to fall from rest at $r=R$ into the surface
of the sphere $r=0$ is 

\begin{equation}
\tau=\frac{(R^{3}+r_{g}^{3})^{\frac{1}{3}}}{2}\left(\frac{R^{3}+r_{g}^{3}}{r_{g}}\right)^{\frac{1}{6}}\left[\arccos(\frac{2r_{g}}{(R^{3}+r_{g}^{3})^{\frac{1}{3}}-1})+\sin\arccos(\frac{2r_{g}}{(R^{3}+r_{g}^{3})^{\frac{1}{3}}-1})\right].\label{eq: tau horizon}\end{equation}

Let us focus the attention on the simplest ball of dust, an interior
that is homogeneous and isotropic everywhere, except at the surface.
This is exactly the case of an interior locally identical to a dust
filled Friedmann closed cosmological model \cite{key-1,key-9}. In
fact, the closed model is the only one of interest because it corresponds
to a gas sphere whose dynamics begins at rest with a finite radius
\cite{key-1,key-9}. The ordinary line-element is given by \cite{key-1,key-9}
\begin{equation}
ds^{2}=-d\tau^{2}+a(\tau)(+d\chi^{2}+\sin^{2}\chi(d\theta^{2}+\sin^{2}\theta d\varphi^{2}),\label{eq: metrica conformemente piatta}\end{equation}

where $a(\tau)$ is the scale factor of the internal space-time. In
the case of zero pressure the stress-energy tensor is 

\begin{equation}
T=\rho u\otimes u,\label{eq: stress energy}\end{equation}

where $\rho$ is the density of the star and $u$ the 4-vector velocity
of the matter. Thus, the Einstein field equations give only one meaningful
relation in terms of $\eta$ \cite{key-1,key-9}

\begin{equation}
(\frac{da}{d\eta})^{2}+a^{2}=\frac{\rho}{3}a^{4},\label{eq: Einstein 1}\end{equation}

which admits the familiar cycloidal solution \cite{key-1,key-9}

\begin{equation}
a=\frac{a_{0}}{2}(1+\cos\eta),\label{eq: cicloide}\end{equation}

and \begin{equation}
\tau=\frac{a_{0}}{2}(\eta+\sin\eta).\label{eq: tau interno}\end{equation}

where $a_{0}$ is a constant. 

Homogeneity and isotropy are broken only at the star's surface which
lies at a radius $\chi=\chi_{0}$ for all $\tau$ during the collapse
\cite{key-1,key-9}, as measured in terms of the co-moving hyper-spherical
polar angle $\chi.$ The match between the internal solution given
by Eqs. (\ref{eq: cicloide}) and (\ref{eq: tau interno}) and the
external solution given by Eqs. (\ref{eq: r}) and (\ref{eq: tau finale})
is possible. As a verification of such a match let us examine the
separate and independent predictions made by the internal and external
solutions for the star's circumference \cite{key-1}. From Eqs. (\ref{eq: r})
and (\ref{eq: tau finale}) the external solution enables the relations: 

\begin{equation}
\begin{array}{c}
C=2\pi(r^{3}+r_{g}^{3})^{\frac{1}{3}}=2\pi(R^{3}+r_{g}^{3})^{\frac{1}{3}}(1+\cos\eta)\\
\\\tau=\frac{(R^{3}+r_{g}^{3})^{\frac{1}{3}}}{2}\left(\frac{R^{3}+r_{g}^{3}}{r_{g}}\right)^{\frac{1}{6}}(\eta+\sin\eta).\end{array}\label{eq: non euclidee}\end{equation}

From Eqs. (\ref{eq: cicloide}) and (\ref{eq: tau interno}) the internal
solution enables the relations: 

\begin{equation}
\begin{array}{c}
C=2\pi(r^{3}+r_{g}^{3})^{\frac{1}{3}}=2\pi\frac{a_{0}\sin\chi_{0}}{2}(1+\cos\eta)\\
\\\tau=\frac{a_{0}}{2}(\eta+\sin\eta).\end{array}\label{eq: non euclidee 2}\end{equation}

Thus, the match works for all time during the collapse if and only
if

\begin{equation}
\begin{array}{c}
R=\left(a_{0}^{3}\sin^{3}\chi_{0}-r_{g}^{3}\right)^{\frac{1}{3}}\\
\\r_{g}=a_{0}\sin^{3}\chi_{0}.\end{array}\label{eq: Match}\end{equation}

By inserting the first of Eqs. (\ref{eq: Match}) in Eq. (\ref{eq: r})
one gets

\begin{equation}
r=\frac{1}{2}\left\{ \left[(a_{0}\sin\chi_{0})(1+\cos\eta)\right]^{3}-r_{g}^{3}\right\} ^{\frac{1}{3}}.\label{eq: r finale}\end{equation}

Eq. (\ref{eq: r finale}) represents the run of the collapse for both
the external and internal solutions for $0\leq\eta\leq\frac{2r_{g}}{(R^{3}+r_{g}^{3})^{\frac{1}{3}}-1}.$
When $\eta=\frac{2r_{g}}{(R^{3}+r_{g}^{3})^{\frac{1}{3}}-1}$ it is
$r=0$ and particles reach the Schwarzschild sphere which is the origin
of the coordinate system. For $\eta>\frac{2r_{g}}{(R^{3}+r_{g}^{3})^{\frac{1}{3}}-1}$
Eq. (\ref{eq: r finale}) represents only the trend of the internal
solution and the $r$ coordinate becomes negative (this is possible
because the origin of the coordinate system is the surface of the
Schwarzschild sphere). The $r$ coordinate reaches a minimum $r=-r_{g}$
for $\eta=\pi.$ Thus, we understand that at this point the collapse
terminates and the star is totally collapsed in a singularity at $r=-r_{g}$.
In other terms, \emph{in the internal geometry all time-like radial
geodesics of the collapsing star terminate after a lapse of finite
proper time in the termination point $r=-r_{g}$ and it is impossible
to extend the internal space-time manifold beyond that termination
point}. Thus, the point $r=-r_{g}$ represents a singularity based
on the rigorous definition by Schmidt \cite{key-43}.

Clearly, as all the particle of the collapsing star fall in the singularity
at $r=-r_{g}$ values of $r>-r_{g}$ do not represent the internal
geometry after the end of the collapse, but they will represent the
external geometry. This implies that the external solution (\ref{eq: sol. gen.fin.}),
i.e. \emph{{}``the} \emph{original Schwarzschild solution''} to
Einstein field equations which has been derived for the first time
by Karl Schwarzschild in \cite{key-2} can be analytically continued
for values of $-r_{g}<r\leq0$ and it results physically equivalent
to the solution (\ref{eq: Hilbert}) that is universally known like
the \char`\"{}\emph{Schwarzschild solution}\char`\"{}. In fact, now
the transformation (\ref{eq: definizione}) can be enabled and the
origin of the coordinate system, $r=0,$ $\theta=0,$ $\varphi=0,$
which is the surface of a sphere having radius $r_{g}$ in the $r,\mbox{ }\theta,\mbox{ }\varphi$
coordinates, results transferred in a non-dimensional material point
$\widehat{r}=0,$ $\theta=0,$ $\varphi=0$ in the $\widehat{r},\mbox{ }\theta,\mbox{ }\varphi$
coordinates. Such a non-dimensional material point corresponds to
the point $r=-r_{g},$ $\theta=0,$ $\varphi=0$ in the original $r,\mbox{ }\theta,\mbox{ }\varphi$
coordinates. 

Then, the authors who claim that \emph{{}``the original Schwarzschild
solution''} leaves no room for the science fiction of the BHs, see
\cite{key-18,key-19}, \cite{key-37} - \cite{key-41}, give the wrong
answer. We realize that the misunderstanding is due to an erroneous
interpretation of the different coordinates. In fact, arches of circumference
appear to be $2\pi r$ if the origin of the coordinate system is a
non-dimensional material point in the core of the BH while they do
not appear to be $2\pi r$, but deformed by the presence of the mass
of the central body $M$ if the origin of the coordinate system is
the surface of the Schwarzschild sphere.

The only way to remove the singularity in the core of a BH within
the classical theory of Einstein's general relativity is changing
the hypotheses which govern the internal geometry of the collapsing
star, following for example the ideas in references from \cite{key-7}
to \cite{key-16}.

\section{Conclusion remarks}

In this paper we clarified a issue on the the debate on {}``\emph{the
original Schwarzschild solution}''. As English translations of Schwarzschild\textquoteright{}s
original paper exist, that paper has become accessible to more people.
A misunderstanding arises from the fact that, historically, the so-called
\char`\"{}\emph{standard Schwarzschild solution}'' (\ref{eq: Hilbert})
was not the original Schwarzschild's work, but it is actually due
to Droste \cite{key-20} and Weyl \cite{key-21}. The solution in
refs. \cite{key-20,key-21} has been ultimately enabled like correct
solution by Hilbert in \cite{key-22}. Based on this, there are authors
who claim that the work of Hilbert was wrong and that Hilbert's mistake
spawned BHs and accuse the community of theoretical physicists to
continue to elaborate on this falsehood, with a hostile shouting down
of any and all voices challenging them \cite{key-18,key-19}, \cite{key-37}
- \cite{key-41}.

With the goal to clarify the issue, we re-analysed \emph{{}``the}
\emph{original Schwarzschild solution''} to Einstein field equations
by showing that such a solution arises from an apparent different
physical hypothesis which assumes arches of circumference to be \emph{not
}$2\pi r$, but deformed by the presence of the mass of the central
body $M.$ This assumption enables the origin of the coordinate system
to be \emph{not} a single point, but a spherical surface having radius
equal to the gravitational radius, i.e. the surface of the Schwarzschild
sphere. The solution works for the external geometry of a spherical
static star and circumnavigates the Birkhoff theorem. After this,
we discussed the simplest case of gravitational collapse, i.e. the
spherical radial collapse of a star with uniform density and zero
pressure, by turning attention to the interior of the collapsing object
and the precise word line that its surface follows in the external
geometry. The result is that the singularity within the totally collapsed
spherical object remains. In fact, a coordinate transform that transfers
the origin of the coordinate system, which is the surface of a sphere
having radius equal to the gravitational radius, in a non-dimensional
material point in the core of the black-hole, re-obtains the solution
re-adapted by Hilbert. Thus, \emph{{}``the} \emph{original Schwarzschild
solution''} \cite{key-3} results physically equivalent to the solution
enabled by Hilbert in \cite{key-22}, i.e. the solution that is universally
known like \char`\"{}\emph{the standard Schwarzschild solution}\char`\"{}.
We conclude that Hilbert was not wrong but they are definitively wrong
the authors who claim that \emph{{}``the} \emph{original Schwarzschild
solution''} implies the non existence of BHs \cite{key-18,key-19},
\cite{key-37} - \cite{key-41}. The misunderstanding is due to an
erroneous interpretation of the different coordinates. In fact, arches
of circumference appear to be $2\pi r$ if the origin of the coordinate
system is a non-dimensional material point in the core of the black-hole,
while they do not appear to be $2\pi r$, but deformed by the presence
of the mass of the central body $M$ if the origin of the coordinate
system is the surface of the Schwarzschild sphere.

Therefore, the only way to remove the singularity in the core of a
BH within the classical theory of Einstein's general relativity is
changing the hypotheses which govern the internal geometry of the
collapsing star, following for example the ideas in references from
\cite{key-7} to \cite{key-16}.

\subsubsection*{Acknowledgements }

I thank Herman Mosquera Cuesta and Jeremy Dunning Davies for interesting
discussions on Black Hole physics. I also thank Chrysostomos for correcting
typos in previous versions of this paper. The Institute for Basic
Research and the International Institute for Theoretical Physics and
Mathematics Einstein-Galilei have to be thanked for supporting this
paper.

\end{document}